\documentclass[review]{elsarticle}

\usepackage{lineno,hyperref}
\usepackage{graphicx}
\usepackage{array}
\usepackage{multirow}
\usepackage{amssymb}
\usepackage{amsmath}
\usepackage{textcomp}
\usepackage[euler]{textgreek}
\usepackage{subcaption}
\usepackage{nicefrac}
\usepackage{array}
\usepackage{booktabs}
\usepackage[margin=1.25in]{geometry}
\usepackage{xcolor}
\usepackage{cancel}
\modulolinenumbers[5]

\journal{Physics of the Dark Universe}

\begin{document}

\begin{frontmatter}

\title{Design of the ALPS\,II Optical System}


\author[1]{M.~Diaz Ortiz}
\author[1]{J.~Gleason}
\author[2]{H.~Grote}
\author[1]{A.~Hallal}
\author[3]{M.~T.~Hartman}
\author[1]{H.~Hollis}
\author[3]{K.-S.~Isleif}
\author[2]{A.~James}
\author[4]{K.~Karan}
\author[1]{T.~Kozlowski}
\author[3]{A.~Lindner}
\author[1]{G.~Messineo}
\author[1]{G.~Mueller}
\author[4]{J.~H.~P\~old}
\author[3]{R.~C.~G.~Smith}
\author[3]{A.~D.~Spector\corref{mycorrespondingauthor}}
\cortext[mycorrespondingauthor]{Corresponding author}
\ead{aaron.spector@desy.de}
\author[1]{D.~B.~Tanner}
\author[3]{L.-W.~Wei}
\author[4]{B.~Willke}
\address[1]{Department of Physics, University of Florida, 32611 Gainesville, Florida, USA}
\address[2]{School of Physics and Astronomy, Cardiff University, CF24 3AA Cardiff, Wales, UK}
\address[3]{Deutsches Elektronen-Synchrotron DESY, Notkestr. 85, 22607 Hamburg, Germany}
\address[4]{Max-Planck-Institut für Gravitationsphysik (Albert-Einstein-Institut) and Leibniz Universität Hannover, 30167 Hannover, Germany}

\begin{abstract}
The Any Light Particle Search II (ALPS\,II) is an experiment currently being built at DESY in Hamburg, Germany, that will use a light-shining-through-a-wall (LSW) approach to search for axion-like particles. ALPS\,II represents a significant step forward for these types of experiments as it will use 24 superconducting dipole magnets, along with dual, high-finesse, 122\,m long optical cavities. 
This paper gives the first comprehensive recipe for the realization of the idea, proposed over 30 years ago, to use optical cavities before and after the wall to increase the power of the regenerated photon signal. The experiment is designed to achieve a sensitivity to the coupling between axion-like particles and photons down to $g_{\alpha\gamma\gamma}=2\times10^{-11}\rm GeV^{-1}$ for masses below 0.1\,meV, more than three orders of magnitude beyond the sensitivity of previous laboratory experiments.  The layout and main components that define ALPS\,II are discussed along with plans for reaching design sensitivity. An accompanying paper (Hallal, et al \cite{HET}) offers a more in-depth description of the heterodyne detection scheme, the first of two independent detection systems that will be implemented in ALPS\,II.
\end{abstract}


\end{frontmatter}


\newcommand{\murm}{%
  \ifmmode
    \mathchoice
        {\hbox{\normalsize\textmu}}
        {\hbox{\normalsize\textmu}}
        {\hbox{\scriptsize\textmu}}
        {\hbox{\tiny\textmu}}%
  \else
    \textmu
  \fi
}

\section{Introduction}

While the standard model has been extremely successful, a variety of phenomena suggest that it is still an incomplete description of our universe.  One example of this is known as the strong CP problem in quantum chromodynamics (QCD), in which a term in the QCD Lagrangian will break CP symmetry if one of it constituent variables has a non zero value \cite{Peccei}. While, as an angle, it appears that this variable should be of order unity, no violations of CP symmetry in the strong force have ever been measured \cite{NeutronDM}. 

The solution proposed by Peccei and Quinn was to promote this variable to a dynamic field associated with a new elementary particle called the axion. This pseudo-Goldstone boson is characterized by the energy scale of the Peccei-Quinn symmetry breaking: the axion mass as well as its coupling strengths to standard model particles is inversely proportional to this energy scale. Experimental results indicate that the energy scale is far beyond the reach of particle accelerators resulting in very light-weight and very weakly coupled axions. These axions would also couple to photons via the following term in the Lagrangian \cite{Peccei,Weinberg:1977ma,Wilczek:1977pj}:
\begin{equation}
\mathcal{L}_{a}=g_{a\gamma\gamma}\phi_{a}\vec{E}\cdot\vec{B}\label{Eq:Lag-1}
\end{equation}
Here $g_{a\gamma\gamma}$ is the axion-photon coupling strength, $\phi_{a}$ is the axion field, the oscillating electric field is given by $\vec{E}$, and $\vec{B}$ represents a static magnetic field.

\begin{figure*}[t]
\centering
  \includegraphics[width=0.75\textwidth]{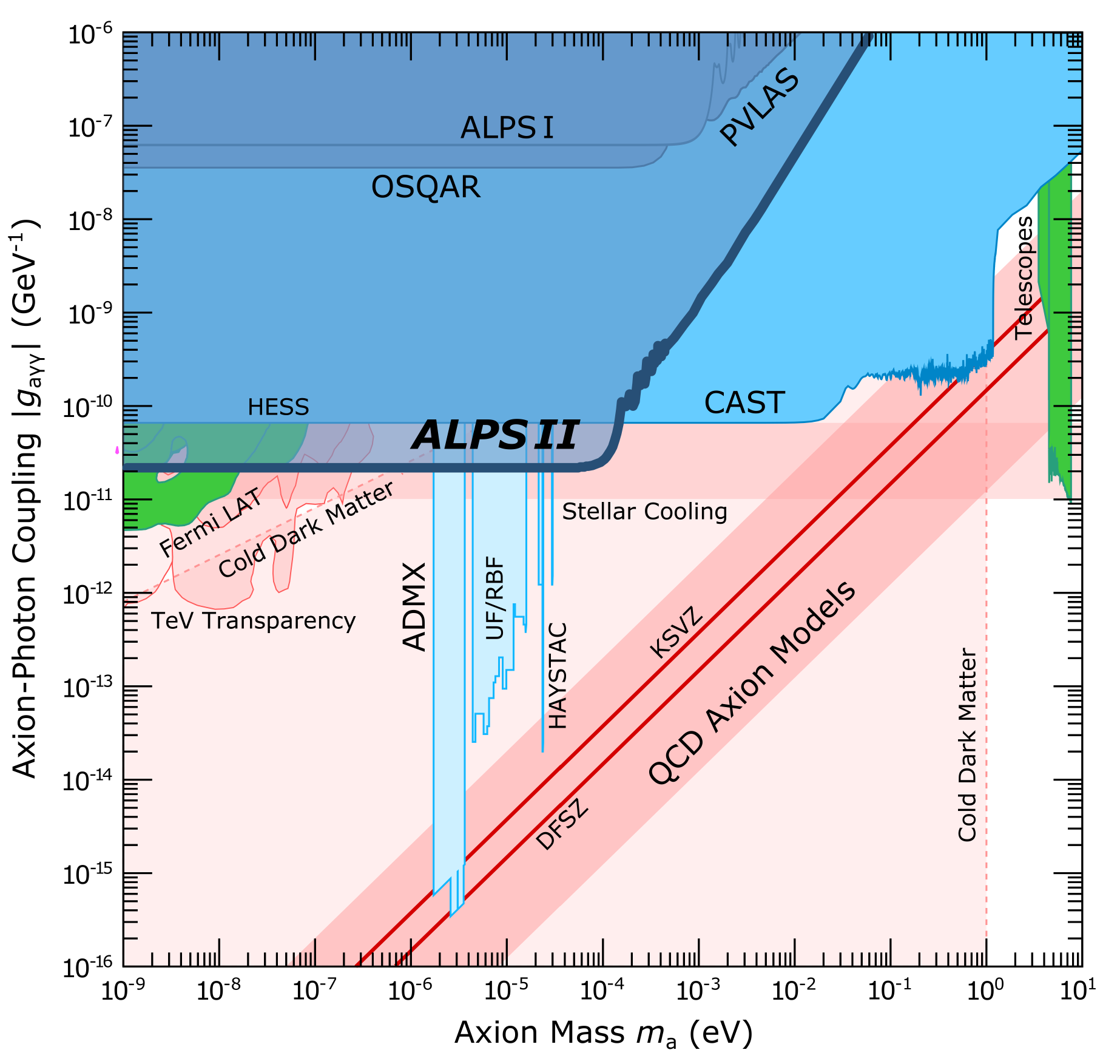}
\caption{ALPS\,II target sensitivity, shown in the transparent blue region, along with current exclusion limits for axion-like-particles given in blue for the helioscopes \cite{CAST}, dark blue for the LSW experiments \cite{ALPSI,OSQAR,PVLASax}, light blue for dark matter haloscopes \cite{ADMX2010,ADMX2018,ADMX2020,ADMXsc,RBF87,RBF89,UF90,YALE,HAYSTAC}, and green for the astronomical observations \cite{HESS,FERMILAT,Telescope}. Regions of the parameter space which offer hints of their existence are shown in red \cite{Kim,SHIFMAN,DINE,ZHIT,CHENG,Meyer,Giannotti,DM1,Dine:1982ah,DM2,DM3}. Please note that several exclusion limits rely on mostly untested assumptions such as axions or axion-like particles making up 100\% of the dark matter in the milky way, structures of galactic and intergalactic magnetic fields, or axion and axion-like particles production in dense stellar plasmas.}\label{Fig:Sens}
\end{figure*}

As it happens, axions could be the cause of a number of other phenomena that the standard model also fails to explain. While the most prominent example of this is the existence of cold dark matter \cite{PDG},  other anomalies such as stellar cooling rates that exceed predictions may also be explained by axions  \cite{Giannotti}. Furthermore, `axion-like' particles which are not themselves the QCD axion, but share similar characteristics to it could explain other anomalous observations such as the transparency of the universe to TeV photons \cite{Meyer}. All of these hints are shown as the red regions in the axion-like particle parameter space in Figure~\ref{Fig:Sens}.

One of the most promising strategies for axion searches involves the Sikivie effect, where in the presence of an external magnetic field, an axion will generate a photon \cite{Sikivie}. There are a number of experiments that try to use this effect to observe axion-like particles and their designs are largely dependent on their respective sources of axions. Haloscopes such as ADMX and MADMAX look for axions that reside in the dark matter halo using microwave resonators immersed in a magnetic field \cite{ADMX2020,MADMAX}. Helioscopes such as CAST and IAXO use the sun as a source of ultra-relativistic axions that then are converted to photons in a magnetic field and detected with X-ray telescopes placed at the end of the magnet \cite{CAST,IAXO}. In contrast to these searches, light-shining-through-a-wall  (LSW) experiments take place entirely in the laboratory using a high-power laser (HPL) propagating through a magnetic field. This generates a beam of axion-like particles that travel through a light-tight wall into a second magnetic field region where some of these axion-like particles convert back to photons \cite{Bibber87}. The current limits on the axion-photon coupling from these different types of searches are shown as the solid blue regions in Figure~\ref{Fig:Sens}, while limits from astronomical observations are shown as the solid green regions.

It is worth emphasizing that LSW experiments make no assumptions regarding the natural prevalence of any of these particles, but merely probe the interactions themselves without the need for an external source. These experiments can therefore determine the photon-coupling strength independent of any astrophysical models, while solar searches and haloscopes, not only depend on the coupling strength, but also rely on models of the axion-flux.

\begin{figure*}[]
\begin{centering}
  \includegraphics[width=\textwidth]{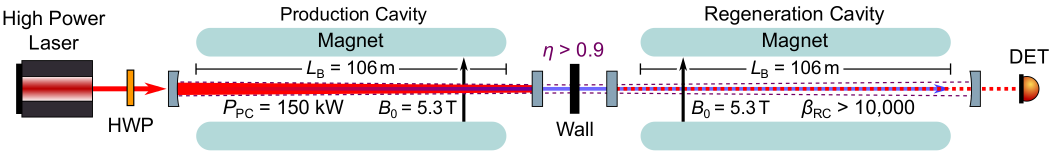}
\caption{Experimental layout of ALPS\,II. The circulating fields in each cavity propagate through 560\,$\rm T\cdot m$ of the  magnetic field times length product, with its length shown as $L_{_{\rm B}}$ and the field strength given by $B_0$. The power inside the PC, shown as solid red, will be at least 150\,kW, while the resonant enhancement of the RC $\beta_{_{\rm RC}}$, must be greater than 10,000. The coupling efficiency between the PC and RC, given by $\eta$, will be at least 0.9, with the projected spatial mode of the PC shown as the dotted purple line. A wall between the cavities prevents any light from the PC from directly entering the RC while a half wave plate (HWP) after the high power laser enables the polarization of the laser to be controlled with respect to the magnetic field. The axion like particle beam is seen as the blue line with the regenerated photon signal represented by the dotted red line that is incident on the detector at the right side of the diagram.}\label{Fig:OSg}
  \end{centering}
\end{figure*}

The Any Light Particle Search II (ALPS\,II) \cite{alpstdr} is a LSW experiment currently under construction at DESY in Hamburg, Germany. With DESY as the host site, the experiment takes advantage of the tunnels, magnets, and cryogenic infrastructure formerly used by the HERA accelerator. It will use 122\,m long optical cavities to amplify light at a wavelength of 1064\,nm that is generated on one side of the wall by a high-power laser (HPL) and by the photon regeneration process on the other side. The experiment is designed to increase the photon regeneration rate by twelve orders of magnitude over earlier LSW experiments \cite{Tanner,Redondo:2010dp,rc,Fukuda}. Like all LSW experiments, ALPS\,II will not only probe the existence of pseudo-scalar fields whose coupling to electromagnetic fields is described by Equation~\ref{Eq:Lag-1}, but also scalar fields whose coupling can be described by the Lagrangian:
\begin{equation}
\mathcal{L}_{s}=g_{a\gamma\gamma}\phi_{s}(\vec{E}^{2}-\vec{B}^{2}).
\label{Eq:LagS-1}
\end{equation}
Experimentally, while the pseudo-scalar fields from Equation~\ref{Eq:Lag-1} require the polarization of the E-field and the B-field to be parallel, scalar fields require that the polarization of the
E-field is orthogonal to the B-field. A signal running in both 
polarization modes with no observed polarization dependence on the 
production rate could be detected as well. This may indicate the existence of other types of Weakly 
Interacting Sub-eV Particles (WISPs) that are produced by kinetic mixing such 
as millicharged particles or hidden sector photons which do not require a static 
magnetic field to interact with photons \cite{Gies}. In the following, we refer to 
particles whose interaction strength depends on $\vec{B}$ as axion-like particles, 
however the observation of any such particle would represent a profound 
discovery as it would be the first detection of an interaction beyond the standard
model.

With a target sensitivity of $2\times10^{-11}{\rm\,GeV^{-1}}$, ALPS\,II aims to go beyond the current CAST limits by a factor of $\sim$3 for masses below 0.1\,meV, investigating a new region of the axion-like particle parameter space that is occupied by hints from both stellar cooling excesses and the TeV transparency. In addition to this, ALPS\,II will also be the first model independent search in a significant region of the parameter space explored by CAST.

To achieve this, the entire optical setup including the two cavities must be tightly controlled to maintain and accurately calibrate the coupling of the generated axion field to the cavities. The following text will focus primarily on the core components that define the optical system for the experiment and discuss how we plan to reach the targeted sensitivity, while also providing a set of top level requirements for the subsystems. A complementary paper \cite{HET}, builds on the work presented here and describes the heterodyne detection scheme (HET) in more detail.

\subsection{ALPS\,II}
\label{Sec:ALP}

\bgroup
\def\arraystretch{1.5}
  \begin{table*}[t]

  \caption{Top level requirements for the first ALPS\,II science run targeting a sensitivity of {$g_{\alpha\gamma\gamma}=2.8\times10^{-11}$\,GeV$^{-1}$} (adapted from \cite{DRD}).\label{tabTLR}}
\begin{tabular*}{\textwidth}{p{1.5cm}  p{13cm}}
\hline\hline
	&	{Requirement} \\
\hline
{TLR1}	&	$>150$\,kW power circulating in PC (fundamental mode, linearly polarized, 1064 nm)	\\
TLR2	&	Parallel and perpendicular polarization adjustment possibility with respect to the magnetic field \\
TLR3	&	Coupling between the axion mode and the RC fundamental mode: $\eta > 90\%$ (power ratio)\\
TLR4	&	RC resonant enhancement $\beta_{_{\rm RC}} \,> \,10\,000$\\	
TLR5 	&	Detector sensitive enough to exclude a reconverted photon rate $>2.8\times10^{-5}$/s with a 95$\%$ confidence level within 20 days \\
TLR6	&	Magnetic field $\times$ length product of 560 T${\cdot}$m for PC and RC magnet string \\
\hline
\end{tabular*}
\end{table*}
\bgroup
\def\arraystretch{1.1}

As mentioned earlier, the optical system for ALPS\,II will consist 
of two 122\,m long, high-finesse optical
cavities whose circulating fields will propagate through strings of
12 superconducting HERA dipole magnets \cite{Magnet}, as shown in
Figure~\ref{Fig:OSg}. A current of 5.7~kA will flow through the
8.8\,m long dipoles and produce a magnetic field of 5.3~T giving
a product of the magnetic field and length of $B_{0}L=560\,\rm T{\cdot}m$ on each 
side of the wall.
On the left side of the wall, light with a wavelength of 1064 nm from the HPL will be resonant with the production cavity (PC), building up the stored energy due to the high cavity finesse.
This light will generate axion-like particles, shown as the transparent blue line in Figure 2, with an identical energy and spatial mode, outlined by the purple dotted line. These particles pass through the
light-tight barrier on the central optical bench (COB) before they
enter the regeneration cavity (RC) where they convert back to photons, shown as the dotted red line.
The regenerated photon rate,
\begin{equation}
n_{{\rm reg}}=\frac{\eta}{16}\left(g_{a\gamma\gamma}F(qL_{_{\rm B}})B_0L_{_{\rm B}}\right)^{4}\frac{P_{_{\rm PC}}}{h\nu}\beta_{_{{\rm RC}}}\label{Eq:Sn},
\end{equation}
scales with $\left(g_{a\gamma\gamma}B_{0}L\right)^{4}$ and is proportional
to the power inside the PC in terms of photons per second, $P_{_{\rm PC}}/h\nu$, and the resonant enhancement $\beta_{_{\rm RC}}$ of the RC \cite{Arias}. The form factor can be 
approximated by the following equation with $L_{_{\rm B}}$ representing the 106\,m 
length of each magnet string.
\begin{equation}
\begin{aligned}
|F(qL_{_{\rm B}})|\approx\left|\frac{2}{qL_{_{\rm B}}}\sin\left(\frac{qL_{_{\rm B}}}{2}\right)\right|  \\ \left(q=\frac{m_{a}^{2}}{2h\nu}\right)
\end{aligned}
\end{equation}
This is a typical phase matching condition which accounts for the possible mass $m_a$, of the relativisitic axion-like particles.
For masses $m_{a}<0.1\,\text{meV}$, the form factor is essentially
unity in ALPS\,II. For a more detailed discussion on this please see \cite{Arias}.

The coupling efficiency $\eta$ between the relativistic
axion field and the eigenmode of the RC takes into account all transversal
and spectral mismatches between the axion mode, which is identical to 
the PC eigenmode, and the eigenmode of the RC. Here $\eta$ is given in terms 
of axion to photon coupling and therefore the axion field to electromagnetic field coupling 
would be given by $\sqrt{\eta}$. It will be possible to verify ${\eta}$ before and 
after measurement runs by opening a shutter in the light tight barrier and allowing the 
PC transmitted field to couple to RC.

Table~\ref{tabTLR} lists the top level requirements (TLR)
of ALPS\,II. While the long magnet string (TLR~6) provides a sensitivity gain of
$\sim25$ in $g_{a\gamma\gamma}$ when compared to ALPS\,I, 
TLR~1, requiring a PC internal power of 150~kW, and TLR~4, 
requiring an RC resonant enhancement $\beta_{_{\rm RC}} \,> \,10\,000$, 
together increase the sensitivity of the experiment by a factor of $\sim40$, demonstrating 
the importance of the optical system.
Achieving both of these requirements depends on
the coatings and surface roughnesses of the cavity mirrors as well
as clipping losses in the magnet strings. This will be discussed further in
section \ref{sec:ALPSII-Cavities}. It should be noted TLR~4 is not far from 
the limits of what is possible for mirrors of these dimensions with 
state of the art polishing techniques. TLR~3 refers to the coupling
of the axion field to the RC and is discussed in section
\ref{sec:Maintaining-the-Axion}. TLR~5 not only establishes requirements 
on the sensitivity of the detection systems, but can also be used to set 
the maximum permissible background signal due to stray light coupling 
from the PC to the detectors.

The first science run will be a search based on the above listed 
parameters that could set an upper limit of 
$g_{\alpha\gamma\gamma}=2.8\times10^{-11}/\text{GeV}$ corresponding 
to an upper limit on the  regenerated
photon rate of $2.8\times10^{-5}/\text{s}$ or roughly 2.4 photons
per 24~h of valid data. This search will be followed by a scalar
particle search at the same sensitivity by changing the polarization
(TLR~2). We will then attempt to improve the sensitivity by increasing the PC
circulating power, the RC resonant enhancement, and the duty cycle to aim
for $g_{\alpha\gamma\gamma}=2\times10^{-11}$/GeV or better \cite{DRD}
for pseudo-scalar and scalar particles.

\subsection{Detection Systems}

ALPS\,II will have the benefit of using two independent detection systems, each with very different systematic uncertainties, to measure the reconverted photons. While two different detector types are not explicitly needed, this approach will help increase confidence that signals observed with the same strength in both detectors, while also above the measured backgrounds, are indeed the result of photon-axion conversion-reconversion process.  The detectors themselves require
different optical systems in order to be operated and therefore cannot be used
in parallel. Working at a wavelength of 1064 nm implies that free-space, off-the-shelf, silicon photodetectors have plenty of sensitivity to be used as detectors in the control system and also for the ALPS II heterodyne readout. Only the single-photon counting approach requires more sensitive detectors, which are under development.

The first detection scheme to be implemented will be the HET, described in an
accompanying paper in this journal \cite{HET}. The HET measures the 
interference beatnote between a laser, referred to as the local oscillator
(LO), and the regenerated photon field on a photodetector. Demodulating
the electronic signal from this photodetector at the known difference
frequency will create a signal proportional to the regenerated field
strength that can be integrated over the measurement time $\tau$.
The regenerated photon signal will thus accumulate proportional to 
$\tau$ while the background signal from laser shot noise will sum 
incoherently and thus be proportional to $\sqrt{\tau}$ \cite{Bush}.

A transition edge sensor (TES) will be used in the
second detection system \cite{Jan}. In contrast to the HET system 
based on interference effects, the TES system will count individual 
photons. The TES consists of an absorptive
tungsten film which is held at a temperature at the threshold of 
superconductivity. When a photon is absorbed by the tungsten it will 
lead to a slight increase in its temperature. This will
suddenly raise the resistance of the chip causing a drop in the bias
current that is flowing through it. This current drop can be
measured over an inductive coil with a superconducting quantum 
interference device. Therefore, the reconverted photons can be 
individually counted as these pulses occur, with an energy resolution 
of $\sim$\,7\%.  A laboratory prototype TES currently triggers at about 0.8\,eV, but this is not necessarily the final configuration. Saturation occurs at energies well above 2.3\,eV (wavelengths below 532\,nm), and can be tuned with the TES working point. A forthcoming paper will offer more details about the TES detection system, including details of the involved optical fiber system as well as the interplay between the 1.17 eV energy of the reconverted photons and the TES system.

\section{ALPS\,II Cavities\label{sec:ALPSII-Cavities}}

Both the PC and RC will be plano-concave cavities with a stability parameter 
$g=0.43$. The curved mirrors will be located at the end stations of the 
experiment while the flat mirrors will be at the center as shown in Figure 
\ref{Fig:OSg}. The radius of curvature of the mirrors at the end stations 
were chosen such that the Rayleigh lengths of the cavity eigenmodes, $z_{R}$, are equal 
to the length of their corresponding magnet strings. This geometry will help minimize aperture 
losses in the cavities while also avoiding higher order mode degeneracies 
that would occur if the cavities were exactly half-confocal. The configuration also 
ensures that the eigenmodes of both cavities can have a high spatial overlap 
as the nominally identical Gaussian beam waists are located on the flat mirrors. 
The distance between the flat cavity mirrors ($<1$\,m) is much smaller than 
the Rayleigh length ($\sim106$\,m) of the modes rendering the resulting loss in 
coupling negligible compared to other contributions.

\begin{table}[t]
\caption{Parameters of the ALPS\,II cavities
\label{Fig:Mir-1}}
\centering
\begin{tabular}{c  c  }
\hline \hline
Parameter (symbol/acronym)	& 	Value 	\\ \hline
Length ($L_{_{\rm PC/RC}}$)		& 	122\,m 	\\
Free spectral range (FSR)	& 	 1.2\,MHz 	\\
Half-width-half-maximum (HWHM) & 	$15\pm2.5\,\text{Hz}$  \\
 End mirror radius of curvature ($R$)&	$214\pm6\,\text{m}$ \\
$1/e^2$ waist radius ($w_0$) 		& 	$6.0\,\text{mm}$  \\
$1/e^2$ end mirror beam radius ($w_{\rm m}$)&	$9.2\,\text{mm}$ \\
Divergence half angle ($\theta_{_{\rm Div}}$) & 57\,{\textmu}rad \\
RC resonant enhancement ($\beta$)&	 $16\,000\pm2\,000$ \\
 \hline
\end{tabular}
\end{table}

The cavity eigenmodes will need to be centered within
the beam tube of the magnet string to reduce clipping losses. The diameter
of the magnet aperture is nominally 55\,mm, however since the magnets were originally used to
steer protons around the arcs of the HERA accelerator, their central axis followed 
a curvature of 600\,m and therefore required straightening. This process was very 
successful, and expanded the free apertures of the magnets from $\sim$37\,mm to 
between 46 and 51\,mm \cite{Magnet}. The largest free aperture magnets
 will be used near the end stations where the beam size and risk of clipping
losses is the highest. 

The magnets along with the rest of the vacuum system that will house 
the cavities are now aligned within $\pm$\,200\,{\textmu}m and $\pm$\,1\,{\textmu}rad of a line 
defining the theoretical optical axis of the experiment. The cavity mirrors will then be placed within 
$\pm$\,1\,mm and $\pm$\,8\,{\textmu}rad of the resulting central line of the 
combined magnet string, reducing clipping losses inside the two cavities to below 1\,ppm.

\subsection{Regeneration cavity}

The resonant enhancement provided by the regeneration cavity:
\begin{equation}
\beta_{_{\rm RC}} \approx\frac{4T_{_{\rm out}}}{\left(T_{_{\rm out}}+T_{_{ b}}+\rho\right)^{2}},
\label{eq:beta}
\end{equation}
is very similar to the power build-up ratio between the input power and the 
maximum circulating power for a generic optical cavity. In this case, it expresses 
the gain in signal power at the detector due to the amplification of the cavity. As 
Equation~\ref{eq:beta} shows, it depends on the losses and transmissivities of each 
mirror. Here $T_{_{\rm out}}$ is the transmissivity of the mirror located nearest to the main 
regenerated photon detector while $T_{_b}$ is the transmissivity of the other cavity mirror. 
Ideally, the maximum resonant enhancement occurs when the RC has minimal round-trip 
losses $\rho$, and the mirror transmissivities are as low as possible and the cavity is in the 
so-called `impedance matched' configuration where $T_{_{\rm out}}=T_{_b}+\rho$.

Losses in our cavities are expected to be dominated by the surface roughness
of the mirrors and the associated scattering of light. Simulations using measured 
surface maps of the actual mirror substrates predict that the scatter losses inside 
the cavity will likely be between 40 and 60\,ppm per round-trip. 
Operating the RC in an `over-coupled configuration' ($T_{_{\rm out}}>T_{_b}+\rho$) 
will simplify the HET detection scheme as it will allow for a more stable power in the 
LO at the science photodetector \cite{HET}.
Therefore, to be conservative, we 
decided to use 100\,ppm as the design value for $T_{_{\rm out}}$ for the initial science run. 
$T_{_b}$ will then be 5\,ppm since the HET also requires some nonzero transmission 
for this mirror to realize the sensing and control scheme. This is discussed in more 
detail in Section~\ref{sec:coupling}.  

The dielectric cavity mirror coatings consist of 
alternating $\lambda/4$ layers of silica 
and tantala to minimize the absorptive losses; note that the same coatings 
will also be used in the PC where laser beam absorption will lead to thermal 
distortions of the cavity eigenmode. These mirrors were received and 
measured to have transmissivities of $\sim110\,\text{ppm}$ and 
$\sim6.7\,\text{ppm}$ at normal incidence which results in an expected resonant enhancement of
\begin{equation}
\beta_{_{\rm RC}}\approx16\,000\,\pm\,2\,000
\end{equation}
given the expected scattering losses and their corresponding uncertainty.

\subsection{Production cavity}

The PC will be seeded with the HPL, a linear polarized laser beam from a single-frequency, low-noise laser system operating at 1064\,nm capable of injecting up to 50\,W to the cavity. The HPL is based on a master-oscillator power-amplifier configuration (MOPA) with an ultra stable Nd:YAG seed laser in non-planar ring-oscillator configuration \cite{HPLB} and a low-noise Nd:YVO$_4$ amplifier \cite{HPLC} with a high spatial purity. The challenge for the laser design is to generate a high power beam with small fluctuations and drifts that allows for a highly efficient coupling of the light into the PC. In this regard the chosen MOPA design is advantageous as it preserves the high frequency stability of the seed laser and has, compared to an oscillator, relatively low intensity levels in the gain material such that thermally induced wave front distortions  and depolarization effects can be minimized.

Light from the HPL will then be incident on the PC. The PC will use mirrors with identical dielectric coatings as the RC mirrors. Therefore, the PC power build-up factor is also expected to be $16\,000\pm2\,000$, the same as the resonant enhancement of the RC. 
A frequency stabilization system with a unity gain frequency of $\sim300\,\text{kHz}$
will maintain the resonance of the laser with respect
to the length of the PC using the standard Pound-Drever-Hall (PDH)
technique \cite{PDH,Black}. The sensing scheme and the loop gain are expected
to keep the laser frequency within a hundredth of the half-width-half-maximum (HWHM) of the
cavity resonance corresponding to a relative power noise inside the
cavity to a RMS value $<$100\,ppm. 

The input optics between the high-power laser (HPL) and the PC will
also be equipped with an automatic alignment system based on a differential wavefront
sensing (DWS) scheme. This system uses a pair of quadrant photodetectors (QPDs)
which measure the lateral shift and angular offset
between the laser mode and the cavity eigenmode \cite{Morrison}.
These signals are then fed back to a pair of actuators to maintain
the alignment of the laser into the cavity. The goal is to also limit the RMS relative power
noise inside the cavity due to alignment fluctuations to $<$100\,ppm
for each degree of freedom. The entire system will guarantee that
the total relative power noise stays below 0.1\% RMS, which should 
be sufficient to reduce the impact of dynamic thermal effects on 
the HET \cite{HET}. The input optics for
the PC will also employ a half-waveplate to
rotate the polarization of the circulating field with respect to the
polarity of the magnet string. This will satisfy TLR2 and allow the
experiment to search for either scalar or pseudo-scalar particles.

This combination of the HPL and cavity
finesse may allow powers as high as 1\,MW inside the PC, however, the final
power level will likely be limited by the absorption in the highly-reflective coating
layers of the two cavity mirrors. There are a number of ways this absorbed 
light could lead to thermal effects that cause higher intracavity losses.
For example, point absorbers heating up on the surface of the mirror could 
cause the formation of low spatial frequency features which, in turn, leads 
to an increase in the scattering losses \cite{Buikema,Glover}. Absorption in the 
mirror coatings could also cause the size of the mode 
circulating in the PC to change and lead to additional clipping loses from the 
beam tube \cite{Winkler}.  The loss in sensitivity due to the mode mismatch between the 
cavity eigenmodes as the PC mirrors heat up is expected to be insignificant in
comparison. Nevertheless, 150\,kW with the beam size on the PC mirror is roughly 
equivalent to the peak circulating intensity in the aLIGO pre-mode cleaner and is 
therefore expected to be attainable \cite{Kwee:12}.

\section{Maintaining the Axion Field Coupling to the RC\label{sec:Maintaining-the-Axion}}
\label{sec:coupling}

  \begin{table}[t]
  \centering
  \caption{Requirements for fulfilling TLR3 \label{tab:tlr3}}
\begin{tabular}{ l  c  }
\hline\hline
Parameter	 (symbol)&  Requirement \\ \hline
\textbf{\textit{Axion coupling to RC }} ($\eta$) &	$>$90\% 		\\
\hspace{0.5cm} \textit{Coherence} ($\eta_{_{\Delta f}}$) 	& 	$>$95\% 	\\
\hspace{1cm} dynamic phase noise ($\Delta\phi_{_{\rm SD}}$)		&	$<$0.2\,rad \\
\hspace{1cm} static frequency offset ($\Delta f$)				&	$<$1.5\,Hz \\
\hspace{0.5cm} \textit{Spatial overlap}	($\eta_{_{\rm T}}$) 		& 	$>$95\% 	\\ 
\hspace{1cm} Angular alignment ($\Delta\theta$)	&  	$<$5.7\,{\textmu}rad \\
\hspace{1cm} Transversal shift ($\Delta x$) 	&	$<$1.2\,mm	\\
\hline
\end{tabular}
\end{table}

The primary obstacles to optimizing the coupling of the axion field to the
RC are related to maintaining its coherence and spatial mode 
matching with respect to the RC eigenmode. These parameters 
will depend on the residual changes of the frequency and 
spatial mode of the PC circulating field with respect to the RC. As 
Table~\ref{tab:tlr3} shows, we allow each to contribute a 5\% loss 
of the signal to meet the 90\% coupling efficiency listed under TLR~3. 
Here we can also see that each of these requirements can be further 
subdivided into requirements on specific parameters of the optical 
system. The following sections discuss how each of these 
requirements are derived and how the optical system will work to satisfy them.

\subsection{Coherence of the PC field with the RC}

Maintaining the coherence between the electromagnetic field regenerated from 
the axion beam and the RC eigenmode is critical to ALPS\,II achieving its target sensitivity.
Therefore, the optical system is designed such that the regenerated field 
should experience no more than 5\% 
average reduction from its optimal resonant enhancement over the 
duration of the measurement due to its frequency noise with respect 
to the RC resonance frequency. This requirement is further divided into one 
on static frequency offset, denoted by $\Delta f$, and one on the dynamic 
phase noise, denoted by $\Delta\phi_{_{\rm SD}}$. As the regenerated 
field is a replica of the field circulating in the PC, the first challenge is to 
accurately tune the frequency of the PC transmitted field to be 
resonant with the RC. The second challenge is to precisely control the 
phase of the PC transmitted field around this nominal value.

\subsubsection{PC Tuning}
\label{sec:pct}

\begin{figure*}[b]
\centering
  \includegraphics[width=\textwidth]{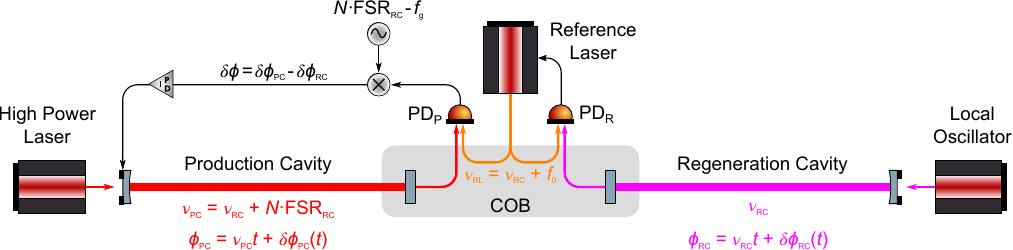}
\caption{Layout of the cavities and control architecture for maintaining the phase lock of the PC in the HET optical system.}\label{Fig:OSb}
\end{figure*}

The signal loss in regenerated photons due to a small
offset of $\Delta f$ in the frequency of the PC transmitted light
relative to a resonance frequency of the RC is quadratic in the offset 
and can be approximated by the following expression.
\begin{equation}
1-\eta_{\Delta f}\approx\left(\frac{\Delta f}{\rm HWHM}\right)^{2}\label{eq:freC-1}
\end{equation}
Here $\rm HWHM$ is the half-width half-max linewidth of the RC. To limit
the loss of regenerated photons to 1\%, we require that the detuning
is less than 10\% of the HWHM or less than 1.5~Hz during the science
run.

As Figures~\ref{Fig:OSb} shows, the optical system when using the HET 
will control the frequency of the PC transmitted field, $\nu_{_{\rm PC}}$, 
via a series of offset phase lock loops (PLLs). This will allow $\nu_{_{\rm PC}}$ 
to track the changes in the resonance frequency 
of the RC due to the presence of environmental noise. To do this the LO
is stabilized to the length of the RC with a high bandwidth control loop 
($>$300\,kHz) ensuring that the frequency of the transmitted 
LO field is equivalent to the RC resonance $\nu_{_{\rm RC}}$. 
The phase of the transmitted LO field, 
$\phi_{_{\rm RC}}$, will therefore be encoded with the RC length noise. 
An auxiliary laser, referred to as the reference laser (RL), will then be phase 
locked to $\nu_{_{\rm RC}}$ with some offset $f_0$ giving it a frequency 
$\nu_{_{\rm RL}} = \nu_{_{\rm RC}} + f_0$. By setting $f_0\neq FSR_{_{\rm RC}}$, 
the RL will not be resonant in the RC. Finally, since the HPL is frequency 
stabilized to the length of the PC, an offset phase lock loop between RL and the 
PC transmitted field will be established to suppress the dynamic phase 
noise $\delta\phi_{_{\rm PC}} - \delta\phi_{_{\rm RC}}$ (as discussed in 
Section~\ref{sec:plpc}). This  gives $\nu_{_{\rm PC}}$ the following value with 
any unintended offsets expressed as $\Delta f$:
\begin{equation}
\nu_{_{\rm PC}} = \nu_{_{\rm RL}} - f_0 + N\cdot FSR_{_{\rm RC}} + \Delta f= \nu_{_{\rm RC}} + N\cdot {\rm FSR_{_{\rm RC}}} + \Delta f .
\end{equation}
With $\Delta f$ sufficiently small, the PC circulating field should be offset 
in frequency from $\nu_{_{\rm RC}}$ by some integer multiple of ${\rm FSR_{_{\rm RC}}}$, $N$, 
thus making it resonant in the 
RC. $\Delta f$ will be minimized tuning the PLL offset frequencies to maximize 
the coupling of the PC transmitted field to the RC with the shutter open. During 
the science runs these values must be maintained when the shutter is closed 
\cite{HET}. This approach requires that the source for the offset frequency and 
the FSR of the RC are both stable. 

The RL-LO beatnote frequency, $f_0$, will be derived from a clock that
is synchronized to a 10\,MHz rubidium frequency standard with a yearly
frequency drift on the order of mHz, well below our requirement. However,
macroscopic changes of the length of the RC will change the optimum
offset frequency by:
\begin{equation}
\Delta f_{_{\rm \Delta FSR}}=N\cdot\Delta {\rm FSR}_{_{\rm RC}}=N\cdot {\rm FSR}_{_{\rm RC}}\frac{\Delta L_{_{\rm RC}}}{L_{_{\rm RC}}}
\end{equation}
The length changes of the RC then have to be
\begin{equation}
\Delta L_{_{\rm RC}}<\frac{\Delta f_{_{\rm \Delta FSR}}}{\rm FSR_{_{\rm RC}}}\frac{L_{_{\rm RC}}}{N}=\frac{150\,\text{{\textmu}m}}{N}
\end{equation}
between retuning measurements to ensure that $\Delta f<1.5\,\text{Hz}$.
A FSR sensing system that uses a modified PDH sensing
technique which uses phase modulated sidebands at some multiple $(\neq N)$
of the FSR will also be implemented \cite{Thorpe}. With this $\Delta L$ will be measured continuously during the science run and if it becomes larger than $150\,\text{{\textmu}m}/N$ the run will be paused and the length of the RC will be adjusted back to its initial value before it is started again. Options to actively control the length of the RC during measurement runs are also being evaluated.

\subsubsection{Phase lock of the PC}
\label{sec:plpc}

The series of PLLs mentioned in the previous section will 
also be used to reduce the frequency or
phase fluctuations $\delta\phi(t)$, of the PC transmitted field relative
to the resonant frequency of the RC. This system must provide 
the precision necessary to meet the requirements on the coherence. 
Phase noise will spread the energy of the ideally monochromatic PC transmitted 
field over a finite frequency band and only the frequency components which 
are resonant in the RC will contribute to the signal. The energy in all frequency 
components outside the line-width of the RC will be attenuated. 

We require that the power integrated over all frequency
components outside this bandwidth is less than 4\% of the total power.
This requirement roughly translates into an upper limit for the standard deviation (SD)
of the phase noise evaluated over the storage time $\mathcal{T}$ of the cavity of \cite{EPJTI}:
\begin{equation}
\Delta\phi_{_{\rm SD}}(t)\approx\sqrt{\left<\delta\phi^{2}(t)\right>_{\mathcal{T}}}<0.2\,\text{rad}
\end{equation}
which the phase lock loop between the PC transmitted field and the
reference field must achieve. 

Due to the high gain and fast bandwidth ($\sim300$\,kHz) of the control loop that stabilizes
the HPL to the PC, the phase of the PC transmitted light should be almost entirely 
determined by the PC length. The PLL must then act on
the length of the PC such that it follows all length changes of the
RC which are impressed on the phase of the reference laser as shown in Figure~\ref{Fig:OSb}.

For this purpose, we developed a piezo-electric actuated mirror mount for the $2"$ 
diameter PC input mirror that supports a control bandwidth of 
4\,kHz. Based on seismic measurements in the HERA tunnel, this bandwidth, paired 
with an aggressive gain function is expected to be sufficient to suppress the 
environmental noise \cite{EPJTI}. 

\subsubsection{Maintaining coherence with the TES optical system}

\begin{figure*}[]
\centering
  \includegraphics[width=\textwidth]{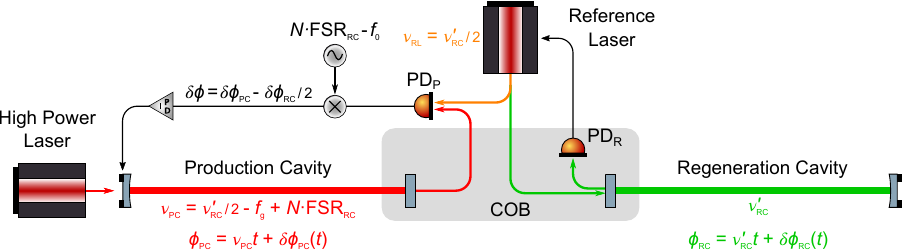}
\caption{Layout of the cavities and control architecture for maintaining the phase lock of the PC in the TES optical system.}\label{Fig:OSt}
\end{figure*}

The TES optical system is also designed to maintain 
the resonance of the PC circulating field with respect to the RC, 
but has slightly different constraints due to the nature of the detector. While 
the HET is capable of distinguishing fields with sub-Hz frequency 
differences \cite{Bush}, the presence of these signals at the detector would 
saturate the TES, in addition to creating backgrounds indistinguishable from the 
regenerated field. Therefore, the frequency of any field that may be incident 
on the TES must be different enough from $\nu_{_{\rm PC}}$ such 
that it can be sufficiently suppressed by optical filters and can also be distinguished 
from the signal field during data processing. For this reason the optical 
system for the TES will appear as seen in Figure~\ref{Fig:OSt}. Here the RL is frequency 
doubled from 1064\,nm to a wavelength of 532\,nm and the laser is then 
frequency stabilized to the RC length with the green light coupled to the cavity. 
With this $\nu_{_{\rm RL}} = \nu'_{_{\rm RC}}/2$, where $\nu'_{_{\rm RC}}$ 
is the resonance frequency of the RC for 532\,nm. This allows the green light to 
be filtered out of the path to the detector (not shown). Furthermore, the TES 
is capable of differentiating between photons with 
wavelengths of 532\,nm and 1064\,nm.

As is the case in the HET optical system, the PC transmitted field is then offset 
phase locked to the 1064\,nm light from RL by actuating on the length of the 
PC. The frequency of the PC transmitted field is therefore:
\begin{equation}
\nu_{_{\rm PC}} = \nu_{_{\rm RL}} - f_{\rm g} + N\cdot FSR_{_{\rm RC}} = \frac{\nu'_{_{\rm RC}}}{2} - f_{\rm g}+ N\cdot FSR_{_{\rm RC}} 
\end{equation}
In this equation $f_{\rm g}$ accounts for the relative frequency offset between 
the green and IR resonances due to the difference in the reflected phase between 
the two wavelengths. To ensure that the PC circulating field is resonant in the RC, 
the beatnote frequency, $\nu_{_{\rm PC}} - \nu_{_{\rm RL}}$, must 
be tuned for either an even or odd number integer multiple of green wavelengths 
in the RC because of $f_{\rm g}$. If the RC length for green is an odd number 
wavelengths of the 532\,nm frequency double light from RL and 
the beatnote frequency is set for an 
even number of wavelengths, then $\nu_{_{\rm PC}}$ will not be resonant with the 
RC length. Like the optical system for HET, this tuning will be optimized with the shutter 
open, by maximizing the power of the PC transmitted light that couples to the RC. 

In Figure~\ref{Fig:OSt} we can see that, like the optical system for the HET, the 
PLL between RL and the light transmitted by the PC is used to control its length 
by feeding back to a piezo-electric actuator at the curved mirror. Other than the 
differences discussed in this section, the optical system for the TES 
adheres to the same requirements and uses the same methods as those that are 
used for the HET optical system and discussed in Sections~\ref{sec:pct} and 
\ref{sec:plpc}.

\subsection{Maintaining Spatial Coupling}

\label{Sec:SO}

\begin{figure*}[t]
\includegraphics[width=1\textwidth]{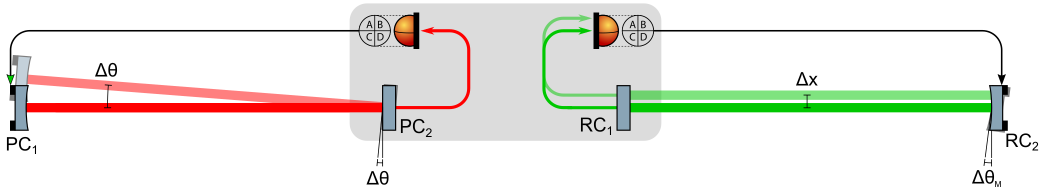}
\caption{Control architecture for maintaining spatial overlap. The QPDs on
the COB monitor the position of the cavity eigenmodes on the cavity
mirrors $\rm PC_2$ and $\rm RC_1$. These signals are used to align the
cavity mirrors at the end stations $\rm PC_1$ and $\rm RC_2$. The aligned configuration is shown as the solid image in the foreground, while examples of the potential alignment errors are shown as transparent images in the background.} 
\label{Fig:OSal}
\end{figure*}

Another effect which could lead to a loss in sensitivity is related to the alignment of the axion field into the RC. Any difference in the angular alignment or transversal position between the incoming axion field and the optical axis of the cavity eigenmode will reduce the coupling of the axion field to the cavity and lead to a decrease in the regenerated photon rate. To be clear, this will not lead to any misalignment between the detection system and the resonant spatial part of the reconverted field that is amplified in the RC, as its spatial mode will be in the eigenmode of the RC, which is itself aligned to the detectors.

The spatial mode of the axion field entering the RC is an extension of the spatial mode of the field inside the PC and the loss in coupling due to small alignment errors can be calculated from the following equation:
\begin{equation}
1-\eta_{_{\rm TM}}\approx\left(\frac{\Delta x}{w_{_{0}}}\right)^2+\left(\frac{\Delta\theta}{\theta_{_{\rm Div}}}\right)^2
\end{equation}
where $\Delta x$ is the transversal shift and $\Delta\theta$ is
the angular misalignment between the two modes measured at the waist
of the RC \cite{UFax,REAPER}. The power loss is quadratic in both terms and required to 
be less than 5\% in total.  With a waist size of $w_0=6$\,mm, our 
requirements of $\Delta x < 1.2$\,mm correspond to a loss in the mode 
matching of 4\% due to transversal shifts. Since the divergence angle 
of the cavities is $\theta_{_{\rm Div}}=57$\,{\textmu}rad, the requirements 
of $\Delta\theta<5.7$\,{\textmu}rad are equivalent to a 1\% loss in the 
mode matching. The systems that will control the alignment of the cavity mirrors 
are discussed in the following sections.

\subsubsection{Central Mirror Alignment}

As the left side of Figure~\ref{Fig:OSal} shows, due to the geometry of the cavities, 
the angular misalignment between their 
eigenmodes will be equal to the relative misalignment of the 
flat mirrors at the center of the experiment. Therefore the alignment of these 
mirrors must be maintained to better than 5.7\,{\textmu}rad to meet our 
requirements. To accomplish this, both mirrors are mounted to the central 
optical bench (COB), which is effectively a large stable surface designed 
for this purpose. It uses no active alignment system and the system instead 
relies on its passive stability. A system for monitoring the relative alignment 
drift between the mirrors on the COB for the HET optical system is detailed 
in \cite{HET}, but not shown in Figure~\ref{Fig:OSal}. 

The COB is constructed from a single aluminum plate
on which all mirrors, beam splitters, and waveplates, as well as position
sensors are either mounted directly or through additional ULE base
plates \cite{HET} using ultra-stable optical mounts.
Tests with an autocollimator have shown that a prototype COB was
capable of maintaining a long term alignment stability of 2\,{\textmu}rad
over one week in air with measured thermal alignment coefficients
of $\approx4$\,{\textmu}rad/K in pitch and $\approx1$\,{\textmu}rad/K
in yaw \cite{BASEPLATE,Wei:20}. The air conditioning system of the cleanroom 
has been designed to maintain an 0.1~K absolute temperature 
stability which would, in principle, eliminate any relevant misalignment.
However, the impact of the heating of mirror $\rm PC_2$ by the cavity internal
field on the alignment still needs to be evaluated during the commissioning
of the experiment. 

\subsubsection{End Mirror Alignment}
\label{Sec:EMa}

Transversal shifts in the relative positions of the eigenmodes will be determined 
by the relative alignment of the curved mirrors with respect to the flat 
mirrors on the COB, ($\Delta \theta_{_{\rm M}}$). In this case, the coupling is 
determined by the ROC of the curved mirrors, $R$.
\begin{equation}
\Delta x = R \Delta \theta_{_{\rm M}}
\end{equation}
We should note here that $\Delta \theta_{_{\rm M}}$ is not the same as the 
misalignment given by the cavity eigenmodes $\Delta \theta$. With a 
radius of curvature in the curved mirrors of 214\,m, the relative alignment 
between the curved and flat mirrors for each cavity must be stabilized to better 
than 5.6\,{\textmu}rad to meet our requirements of $\Delta x<1.2$\,mm. As 
we can see, this is quite similar to the requirements on the alignment of the 
flat mirrors on the COB.

As Figure~\ref{Fig:OSal} shows, this is done using
two in-vacuum QPDs housed on the COB to monitor
the position of the cavity eigenmodes by performing
DC differential measurements of the light incident on their quandrants.
These QPDs are optimized to sense the position of the 6\,mm radius
beam with sub 100\,{\textmu}m precision\,\cite{Wei:20}. Assuming that the 
components on the COB remain stationary, the positions of the
eigenmodes on the flat cavity mirrors $\rm PC_2$ and $\rm RC_1$, and
in extension on the two QPDs, only depend on the orientations of the
curved cavity mirrors $\rm PC_1$ and $\rm RC_2$, respectively. The differential
signals from the QPDs will be fed back to active alignment stages
that are capable of controlling the pitch and yaw of the curved cavity
end mirrors. 

As we will see in the next section, due to the presence of wedge angles in 
the optics in between the cavities mirrors on the COB, the eigenmodes will
have some static transversal shift between them. Nevertheless, this is 
expected to be below 200\,{\textmu}m. When including the
 $<3$\,{\textmu}rad offsets in the COB cavity mirrors the cumulative mode 
matching losses due to the static misalignments should be:
\begin{equation}
1-\eta_{_{\rm TM}}\approx\left(\frac{\Delta x}{w_{_{0}}}\right)^2+\left(\frac{\Delta\theta}{\theta_{_{{\rm Div}}}}\right)^2<0.5\%.
\end{equation}
This is on the same order as the 
matching between the axion field and the PC transmitted field after 
traversing the COB optics, leaving significant 
margin for systematic errors and drifts. 

\subsubsection{Quantifying the transversal matching}

Like the tuning of PLL offset frequency, the alignment of the cavity 
eigenmodes will be quantified using the PC transmitted field when
the shutter is open. This quantification has a systematic error due
to the refraction of the beams that pass through the optical 
components located between the cavity
internal fields. The substrates of the cavity end mirrors have known
wedge angles $\theta_{_{\rm W}}$ between $3$ and 4\,{\textmu}rad which will refract
the PC transmitted field, but not the axion field. By clocking the
two substrates correctly to minimize their differential wedge angle 
$\Delta\theta_{_{\rm W}}$, the refraction angles will
compensate each other such that the final angular refraction is:
\begin{equation}
\theta_{_{\rm refr}}=(n-1)\Delta\theta_{_{\rm W}}<1\,\text{{\textmu}rad}
\end{equation}
The refraction in $\rm PC_2$ will laterally shift the PC transmitted
beam by $\Delta x<4\,${\textmu}m, well below the level of the 
other sources of alignment noise. 

The optics in between the cavity mirrors are also made from
substrates with known wedge angles between 2 and 5\,{\textmu}rad
and will also be clocked to reduce the overall deflection to below
2\,{\textmu}rad. The total deflection angle between the beams
will therefore be below 3\,{\textmu}rad.

Each substrate in between the cavity mirrors will laterally shift the 
beam off of the optical axis of the system by:
\begin{equation}
y=d\frac{\sin(\theta_{1}-\theta_{2})}{\cos\theta_{2}}\qquad\sin\theta_{1}=n\sin\theta_{2}
\end{equation}
where $d$ is the thickness of the substrate, $\theta_{1}$ the angle
of incidence, $\theta_{2}$ the angle inside the material, and $n$ the index 
of refraction. To cancel the cumulative effect, the number of substrates 
which shift the beam to the left has to be equal to the number of substrates 
which shift the beam to the right, assuming that the substrates are equal 
in thickness and in material, and that the angle of incidence is the same. 
The COB design \cite{HET} uses two substrates
that shift the beam left and two that shift it right; all at $35^{\circ}$
angle of incidence and all made from fused silica. According to the
vendor, the substrates are 9.5\,mm thick with a tolerance of $[+0,-0.5\,\text{mm}]$. This results
in a worst case lateral shift of $\sim200$\,{\textmu}m. 

Based on these numbers, the uncertainty in the resulting mode mismatch between the PC
transmitted field and the axion field for both degrees of freedom
will be below 0.4\% which will allow us to use the PC transmitted field
to verify that the overall alignment of the axion mode into the RC is better 
than 95\%.

\section{Summary and Conclusion}

ALPS\,II is designed to be the most sensitive LSW experiment to date with a regenerated photon rate more than 12 orders of magnitude larger than previous experiments. The innovations in the optical system alone account for more than 6 orders of magnitude of the increase in the regenerated signal while the additional boost will come from the long magnetic field length and improvements in detector technologies \cite{Magnet,Jan,HET}. Two 122\,m, high-finesse optical cavities are largely responsible for the gains from optical system. The use of these cavities, however, presents a unique set of challenges related to maintaining their coherence and spatial overlap. In this paper we have described the core components of ALPS\,II, and how the optical system is designed to address these challenges.

We will take great care that the spatial mode of the transmitted light from the PC that is incident on the RC is an accurate representation of the axion mode. This will allow us to quantify the coherence and mode matching of the axion field with respect to the RC. The length and alignment sensing system for the lasers and the cavities is based on PDH and DWS, well established phase sensing schemes with sufficient sensitivity to monitor all relevant degrees of freedom. Additionally, we developed and tested different actuators which should have enough range and bandwidth to operate ALPS\,II in the HERA tunnels. 

We also discussed our plans to employ two different schemes to detect the regenerated photon signal. The first one is the HET and is described in detail in an accompanying paper, while the second scheme uses a TES and will be implemented following the HET science runs. The experiment itself is presently under construction and aiming for a first science run in 2021. Once fully operational, the optical system should allow  ALPS\,II to be able to detect axions with a $95\%$ confidence level for coupling constants as low as $g_{a\gamma\gamma} = 2\times 10^{-11} /\rm{GeV}$ using 20 days of valid science data. 

\section*{Acknowledgments}

The work is supported by the Deutsche Forschungsgemeinschaft [grant number WI 1643/2-1], by the German Volkswagen Stiftung, the National Science Foundation [grant numbers PHY-2110705 and PHY-1802006], the Heising Simons Foundation [grant numbers 2015-154 and 2020-1841], and the UK Science and Technologies Facilities Council [grant number ST/T006331/1].


\bibliography{ALPS_design.bib}

\begin{thebibliography}{10}
\expandafter\ifx\csname url\endcsname\relax
  \def\url#1{\texttt{#1}}\fi
\expandafter\ifx\csname urlprefix\endcsname\relax\def\urlprefix{URL }\fi
\expandafter\ifx\csname href\endcsname\relax
  \def\href#1#2{#2} \def\path#1{#1}\fi

\bibitem{HET}
A.~Hallal, G.~Messineo, M.~D. Ortiz, J.~Gleason, H.~Hollis, D.~Tanner,
  G.~Mueller, A.~Spector,
  \href{https://www.sciencedirect.com/science/article/pii/S2212686421001382}{The
  heterodyne sensing system for the {ALPS\,II} search for sub-ev weakly
  interacting particles}, Physics of the Dark Universe (2021) 100914\href
  {https://doi.org/https://doi.org/10.1016/j.dark.2021.100914}
  {\path{doi:https://doi.org/10.1016/j.dark.2021.100914}}.
\newline\urlprefix\url{https://www.sciencedirect.com/science/article/pii/S2212686421001382}

\bibitem{Peccei}
R.~D. Peccei, H.~R. Quinn, {CP} conservation in the presence of
  pseudoparticles, Phys. Rev. Lett. 38~(25) (1977) 1440.

\bibitem{NeutronDM}
C.~Abel, S.~Afach, N.~J. Ayres, C.~A. Baker, G.~Ban, G.~Bison, K.~Bodek,
  V.~Bondar, M.~Burghoff, E.~Chanel, Z.~Chowdhuri, P.-J. Chiu, B.~Clement,
  C.~B. Crawford, M.~Daum, S.~Emmenegger, L.~Ferraris-Bouchez, M.~Fertl,
  P.~Flaux, B.~Franke, A.~Fratangelo, P.~Geltenbort, K.~Green, W.~C. Griffith,
  M.~van~der Grinten, Z.~D. Gruji\ifmmode~\acute{c}\else \'{c}\fi{}, P.~G.
  Harris, L.~Hayen, W.~Heil, R.~Henneck, V.~H\'elaine, N.~Hild, Z.~Hodge,
  M.~Horras, P.~Iaydjiev, S.~N. Ivanov, M.~Kasprzak, Y.~Kermaidic, K.~Kirch,
  A.~Knecht, P.~Knowles, H.-C. Koch, P.~A. Koss, S.~Komposch, A.~Kozela,
  A.~Kraft, J.~Krempel, M.~Ku\ifmmode~\acute{z}\else \'{z}\fi{}niak, B.~Lauss,
  T.~Lefort, Y.~Lemi\`ere, A.~Leredde, P.~Mohanmurthy, A.~Mtchedlishvili,
  M.~Musgrave, O.~Naviliat-Cuncic, D.~Pais, F.~M. Piegsa, E.~Pierre, G.~Pignol,
  C.~Plonka-Spehr, P.~N. Prashanth, G.~Qu\'em\'ener, M.~Rawlik, D.~Rebreyend,
  I.~Rien\"acker, D.~Ries, S.~Roccia, G.~Rogel, D.~Rozpedzik, A.~Schnabel,
  P.~Schmidt-Wellenburg, N.~Severijns, D.~Shiers, R.~Tavakoli~Dinani, J.~A.
  Thorne, R.~Virot, J.~Voigt, A.~Weis, E.~Wursten, G.~Wyszynski, J.~Zejma,
  J.~Zenner, G.~Zsigmond,
  \href{https://link.aps.org/doi/10.1103/PhysRevLett.124.081803}{Measurement of
  the {Permanent} {Electric} {Dipole} {Moment} of the {Neutron}}, Phys. Rev.
  Lett. 124 (2020) 081803.
\newblock \href {https://doi.org/10.1103/PhysRevLett.124.081803}
  {\path{doi:10.1103/PhysRevLett.124.081803}}.
\newline\urlprefix\url{https://link.aps.org/doi/10.1103/PhysRevLett.124.081803}

\bibitem{Weinberg:1977ma}
S.~Weinberg, A new light boson?, Phys. Rev. Lett. 40~(4) (1978) 223.

\bibitem{Wilczek:1977pj}
F.~Wilczek, Problem of {Strong} {P} and {T} {Invariance} in the {Presence} of
  {Instantons}, Phys. Rev. Lett. 40~(5) (1978) 279.

\bibitem{CAST}
V.~Anastassopoulos, S.~Aune, K.~Barth, A.~Belov, H.~Br{\"a}uninger,
  G.~Cantatore, J.~Carmona, J.~Castel, S.~Cetin, F.~Christensen, et~al., New
  {CAST} limit on the axion--photon interaction, Nat. Phys. 13~(6) (2017) 584.

\bibitem{ALPSI}
K.~Ehret, M.~Frede, S.~Ghazaryan, M.~Hildebrandt, E.-A. Knabbe, D.~Kracht,
  A.~Lindner, J.~List, T.~Meier, N.~Meyer, et~al., New {ALPS} results on
  hidden-sector lightweights, Phys. Lett. 689~(4-5) (2010) 149--155.

\bibitem{OSQAR}
R.~Ballou, G.~Deferne, M.~Finger~Jr, M.~Finger, L.~Flekova, J.~Hosek, S.~Kunc,
  K.~Macuchova, K.~Meissner, P.~Pugnat, et~al., New exclusion limits on scalar
  and pseudoscalar axionlike particles from light shining through a wall, Phys.
  Rev. D 92~(9) (2015) 092002.

\bibitem{PVLASax}
F.~Della~Valle, E.~Milotti, A.~Ejlli, G.~Messineo, L.~Piemontese, G.~Zavattini,
  U.~Gastaldi, R.~Pengo, G.~Ruoso,
  \href{https://link.aps.org/doi/10.1103/PhysRevD.90.092003}{First results from
  the new {PVLAS} apparatus: {A} new limit on vacuum magnetic birefringence},
  Phys. Rev. D 90 (2014) 092003.
\newblock \href {https://doi.org/10.1103/PhysRevD.90.092003}
  {\path{doi:10.1103/PhysRevD.90.092003}}.
\newline\urlprefix\url{https://link.aps.org/doi/10.1103/PhysRevD.90.092003}

\bibitem{ADMX2010}
S.~Asztalos, E.~Daw, H.~Peng, L.~J. Rosenberg, C.~Hagmann, D.~Kinion,
  W.~Stoeffl, K.~van Bibber, P.~Sikivie, N.~S. Sullivan, D.~B. Tanner,
  F.~Nezrick, M.~S. Turner, D.~M. Moltz, J.~Powell, M.-O. Andr\'e, J.~Clarke,
  M.~M\"uck, R.~F. Bradley,
  \href{https://link.aps.org/doi/10.1103/PhysRevD.64.092003}{Large-scale
  microwave cavity search for dark-matter axions}, Phys. Rev. D 64 (2001)
  092003.
\newblock \href {https://doi.org/10.1103/PhysRevD.64.092003}
  {\path{doi:10.1103/PhysRevD.64.092003}}.
\newline\urlprefix\url{https://link.aps.org/doi/10.1103/PhysRevD.64.092003}

\bibitem{ADMX2018}
N.~Du, N.~Force, R.~Khatiwada, E.~Lentz, R.~Ottens, L.~Rosenberg, G.~Rybka,
  G.~Carosi, N.~Woollett, D.~Bowring, et~al., Search for invisible axion dark
  matter with the axion dark matter experiment, Phys. Rev. Lett. 120~(15)
  (2018) 151301.

\bibitem{ADMX2020}
T.~Braine, R.~Cervantes, N.~Crisosto, N.~Du, S.~Kimes, L.~J. Rosenberg,
  G.~Rybka, J.~Yang, D.~Bowring, A.~S. Chou, R.~Khatiwada, A.~Sonnenschein,
  W.~Wester, G.~Carosi, N.~Woollett, L.~D. Duffy, R.~Bradley, C.~Boutan,
  M.~Jones, B.~H. LaRoque, N.~S. Oblath, M.~S. Taubman, J.~Clarke, A.~Dove,
  A.~Eddins, S.~R. O'Kelley, S.~Nawaz, I.~Siddiqi, N.~Stevenson, A.~Agrawal,
  A.~V. Dixit, J.~R. Gleason, S.~Jois, P.~Sikivie, J.~A. Solomon, N.~S.
  Sullivan, D.~B. Tanner, E.~Lentz, E.~J. Daw, J.~H. Buckley, P.~M. Harrington,
  E.~A. Henriksen, K.~W. Murch,
  \href{https://link.aps.org/doi/10.1103/PhysRevLett.124.101303}{Extended
  {Search} for the {Invisible} {Axion} with the {Axion} {Dark} {Matter}
  {Experiment}}, Phys. Rev. Lett. 124 (2020) 101303.
\newblock \href {https://doi.org/10.1103/PhysRevLett.124.101303}
  {\path{doi:10.1103/PhysRevLett.124.101303}}.
\newline\urlprefix\url{https://link.aps.org/doi/10.1103/PhysRevLett.124.101303}

\bibitem{ADMXsc}
C.~Boutan, M.~Jones, B.~H. LaRoque, N.~S. Oblath, R.~Cervantes, N.~Du,
  N.~Force, S.~Kimes, R.~Ottens, L.~J. Rosenberg, G.~Rybka, J.~Yang, G.~Carosi,
  N.~Woollett, D.~Bowring, A.~S. Chou, R.~Khatiwada, A.~Sonnenschein,
  W.~Wester, R.~Bradley, E.~J. Daw, A.~Agrawal, A.~V. Dixit, J.~Clarke, S.~R.
  O'Kelley, N.~Crisosto, J.~R. Gleason, S.~Jois, P.~Sikivie, I.~Stern, N.~S.
  Sullivan, D.~B. Tanner, P.~M. Harrington, E.~Lentz,
  \href{https://link.aps.org/doi/10.1103/PhysRevLett.121.261302}{Piezoelectrically
  {Tuned} {Multimode} {Cavity} {Search} for {Axion} {Dark} {Matter}}, Phys.
  Rev. Lett. 121 (2018) 261302.
\newblock \href {https://doi.org/10.1103/PhysRevLett.121.261302}
  {\path{doi:10.1103/PhysRevLett.121.261302}}.
\newline\urlprefix\url{https://link.aps.org/doi/10.1103/PhysRevLett.121.261302}

\bibitem{RBF87}
S.~DePanfilis, A.~C. Melissinos, B.~E. Moskowitz, J.~T. Rogers, Y.~K.
  Semertzidis, W.~U. Wuensch, H.~J. Halama, A.~G. Prodell, W.~B. Fowler, F.~A.
  Nezrick, \href{https://link.aps.org/doi/10.1103/PhysRevLett.59.839}{Limits on
  the abundance and coupling of cosmic axions at $4.5<{m}_{\rm a}<5.0$
  {\textmu}e{V}}, Phys. Rev. Lett. 59 (1987) 839--842.
\newblock \href {https://doi.org/10.1103/PhysRevLett.59.839}
  {\path{doi:10.1103/PhysRevLett.59.839}}.
\newline\urlprefix\url{https://link.aps.org/doi/10.1103/PhysRevLett.59.839}

\bibitem{RBF89}
W.~U. Wuensch, S.~De~Panfilis-Wuensch, Y.~K. Semertzidis, J.~T. Rogers, A.~C.
  Melissinos, H.~J. Halama, B.~E. Moskowitz, A.~G. Prodell, W.~B. Fowler, F.~A.
  Nezrick, \href{https://link.aps.org/doi/10.1103/PhysRevD.40.3153}{Results of
  a laboratory search for cosmic axions and other weakly coupled light
  particles}, Phys. Rev. D 40 (1989) 3153--3167.
\newblock \href {https://doi.org/10.1103/PhysRevD.40.3153}
  {\path{doi:10.1103/PhysRevD.40.3153}}.
\newline\urlprefix\url{https://link.aps.org/doi/10.1103/PhysRevD.40.3153}

\bibitem{UF90}
C.~Hagmann, P.~Sikivie, N.~S. Sullivan, D.~B. Tanner,
  \href{https://link.aps.org/doi/10.1103/PhysRevD.42.1297}{Results from a
  search for cosmic axions}, Phys. Rev. D 42 (1990) 1297--1300.
\newblock \href {https://doi.org/10.1103/PhysRevD.42.1297}
  {\path{doi:10.1103/PhysRevD.42.1297}}.
\newline\urlprefix\url{https://link.aps.org/doi/10.1103/PhysRevD.42.1297}

\bibitem{YALE}
B.~M. Brubaker, L.~Zhong, Y.~V. Gurevich, S.~B. Cahn, S.~K. Lamoreaux,
  M.~Simanovskaia, J.~R. Root, S.~M. Lewis, S.~Al~Kenany, K.~M. Backes,
  I.~Urdinaran, N.~M. Rapidis, T.~M. Shokair, K.~A. van Bibber, D.~A. Palken,
  M.~Malnou, W.~F. Kindel, M.~A. Anil, K.~W. Lehnert, G.~Carosi,
  \href{https://link.aps.org/doi/10.1103/PhysRevLett.118.061302}{First
  {Results} from a {Microwave} {Cavity} {Axion} {Search} at 24\,{\textmu}e{V}},
  Phys. Rev. Lett. 118 (2017) 061302.
\newblock \href {https://doi.org/10.1103/PhysRevLett.118.061302}
  {\path{doi:10.1103/PhysRevLett.118.061302}}.
\newline\urlprefix\url{https://link.aps.org/doi/10.1103/PhysRevLett.118.061302}

\bibitem{HAYSTAC}
L.~Zhong, S.~Al~Kenany, K.~M. Backes, B.~M. Brubaker, S.~B. Cahn, G.~Carosi,
  Y.~V. Gurevich, W.~F. Kindel, S.~K. Lamoreaux, K.~W. Lehnert, S.~M. Lewis,
  M.~Malnou, R.~H. Maruyama, D.~A. Palken, N.~M. Rapidis, J.~R. Root,
  M.~Simanovskaia, T.~M. Shokair, D.~H. Speller, I.~Urdinaran, K.~A. van
  Bibber, \href{https://link.aps.org/doi/10.1103/PhysRevD.97.092001}{Results
  from phase 1 of the {HAYSTAC} microwave cavity axion experiment}, Phys. Rev.
  D 97 (2018) 092001.
\newblock \href {https://doi.org/10.1103/PhysRevD.97.092001}
  {\path{doi:10.1103/PhysRevD.97.092001}}.
\newline\urlprefix\url{https://link.aps.org/doi/10.1103/PhysRevD.97.092001}

\bibitem{HESS}
M.~V. Beznogov, E.~Rrapaj, D.~Page, S.~Reddy,
  \href{https://link.aps.org/doi/10.1103/PhysRevC.98.035802}{Constraints on
  axion-like particles and nucleon pairing in dense matter from the hot neutron
  star in {HESS J1731-347}}, Phys. Rev. C 98 (2018) 035802.
\newblock \href {https://doi.org/10.1103/PhysRevC.98.035802}
  {\path{doi:10.1103/PhysRevC.98.035802}}.
\newline\urlprefix\url{https://link.aps.org/doi/10.1103/PhysRevC.98.035802}

\bibitem{FERMILAT}
M.~Meyer, M.~Giannotti, A.~Mirizzi, J.~Conrad, M.~A. S\'anchez-Conde,
  \href{https://link.aps.org/doi/10.1103/PhysRevLett.118.011103}{{Fermi}
  {Large} {Area} {Telescope} as a {Galactic} {Supernovae} {Axionscope}}, Phys.
  Rev. Lett. 118 (2017) 011103.
\newblock \href {https://doi.org/10.1103/PhysRevLett.118.011103}
  {\path{doi:10.1103/PhysRevLett.118.011103}}.
\newline\urlprefix\url{https://link.aps.org/doi/10.1103/PhysRevLett.118.011103}

\bibitem{Telescope}
D.~Grin, G.~Covone, J.-P. Kneib, M.~Kamionkowski, A.~Blain, E.~Jullo,
  \href{https://link.aps.org/doi/10.1103/PhysRevD.75.105018}{Telescope search
  for decaying relic axions}, Phys. Rev. D 75 (2007) 105018.
\newblock \href {https://doi.org/10.1103/PhysRevD.75.105018}
  {\path{doi:10.1103/PhysRevD.75.105018}}.
\newline\urlprefix\url{https://link.aps.org/doi/10.1103/PhysRevD.75.105018}

\bibitem{Kim}
J.~E. Kim,
  \href{https://link.aps.org/doi/10.1103/PhysRevLett.43.103}{{Weak-Interaction}
  {Singlet} and {Strong} $\mathrm{CP}$ {Invariance}}, Phys. Rev. Lett. 43
  (1979) 103--107.
\newblock \href {https://doi.org/10.1103/PhysRevLett.43.103}
  {\path{doi:10.1103/PhysRevLett.43.103}}.
\newline\urlprefix\url{https://link.aps.org/doi/10.1103/PhysRevLett.43.103}

\bibitem{SHIFMAN}
M.~Shifman, A.~Vainshtein, V.~Zakharov,
  \href{http://www.sciencedirect.com/science/article/pii/0550321380902096}{Can
  confinement ensure natural {CP} invariance of strong interactions?}, Nuclear
  Physics B 166~(3) (1980) 493 -- 506.
\newblock \href {https://doi.org/https://doi.org/10.1016/0550-3213(80)90209-6}
  {\path{doi:https://doi.org/10.1016/0550-3213(80)90209-6}}.
\newline\urlprefix\url{http://www.sciencedirect.com/science/article/pii/0550321380902096}

\bibitem{DINE}
M.~Dine, W.~Fischler, M.~Srednicki,
  \href{http://www.sciencedirect.com/science/article/pii/0370269381905906}{A
  simple solution to the strong {CP} problem with a harmless axion}, Physics
  Letters B 104~(3) (1981) 199 -- 202.
\newblock \href {https://doi.org/https://doi.org/10.1016/0370-2693(81)90590-6}
  {\path{doi:https://doi.org/10.1016/0370-2693(81)90590-6}}.
\newline\urlprefix\url{http://www.sciencedirect.com/science/article/pii/0370269381905906}

\bibitem{ZHIT}
A.~Zhitnitsky, Sov. J. Nucl. Phys. 31 (1980).

\bibitem{CHENG}
S.~L. Cheng, C.~Q. Geng, W.-T. Ni,
  \href{https://link.aps.org/doi/10.1103/PhysRevD.52.3132}{Axion-photon
  couplings in invisible axion models}, Phys. Rev. D 52 (1995) 3132--3135.
\newblock \href {https://doi.org/10.1103/PhysRevD.52.3132}
  {\path{doi:10.1103/PhysRevD.52.3132}}.
\newline\urlprefix\url{https://link.aps.org/doi/10.1103/PhysRevD.52.3132}

\bibitem{Meyer}
M.~Meyer, D.~Horns, M.~Raue, First lower limits on the photon-axion-like
  particle coupling from very high energy gamma-ray observations, Phys. Rev. D
  87~(3) (2013) 035027.

\bibitem{Giannotti}
M.~Giannotti, I.~Irastorza, J.~Redondo, A.~Ringwald, Cool {WISP}s for stellar
  cooling excesses, J. Cosmol. Astropart. P. 2016~(05) (2016) 057.

\bibitem{DM1}
L.~Abbott, P.~Sikivie,
  \href{http://www.sciencedirect.com/science/article/pii/037026938390638X}{A
  cosmological bound on the invisible axion}, Physics Letters B 120~(1) (1983)
  133 -- 136.
\newblock \href {https://doi.org/https://doi.org/10.1016/0370-2693(83)90638-X}
  {\path{doi:https://doi.org/10.1016/0370-2693(83)90638-X}}.
\newline\urlprefix\url{http://www.sciencedirect.com/science/article/pii/037026938390638X}

\bibitem{Dine:1982ah}
M.~Dine, W.~Fischler, {The {Not} {So} {Harmless} {Axion}}, Phys. Lett. B 120
  (1983) 137--141.
\newblock \href {https://doi.org/10.1016/0370-2693(83)90639-1}
  {\path{doi:10.1016/0370-2693(83)90639-1}}.

\bibitem{DM2}
J.~Preskill, M.~B. Wise, F.~Wilczek,
  \href{http://www.sciencedirect.com/science/article/pii/0370269383906378}{Cosmology
  of the invisible axion}, Physics Letters B 120~(1) (1983) 127 -- 132.
\newblock \href {https://doi.org/https://doi.org/10.1016/0370-2693(83)90637-8}
  {\path{doi:https://doi.org/10.1016/0370-2693(83)90637-8}}.
\newline\urlprefix\url{http://www.sciencedirect.com/science/article/pii/0370269383906378}

\bibitem{DM3}
J.~Ipser, P.~Sikivie,
  \href{https://link.aps.org/doi/10.1103/PhysRevLett.50.925}{Can {Galactic}
  {Halos} {Be} {Made} of{ Axions}?}, Phys. Rev. Lett. 50 (1983) 925--927.
\newblock \href {https://doi.org/10.1103/PhysRevLett.50.925}
  {\path{doi:10.1103/PhysRevLett.50.925}}.
\newline\urlprefix\url{https://link.aps.org/doi/10.1103/PhysRevLett.50.925}

\bibitem{PDG}
M.~Tanabashi, K.~Hagiwara, K.~Hikasa, K.~Nakamura, Y.~Sumino, F.~Takahashi,
  J.~Tanaka, K.~Agashe, G.~Aielli, C.~Amsler, et~al., Review of particle
  physics, Phys. Rev. D 98~(3) (2018) 030001.

\bibitem{Sikivie}
P.~Sikivie,
  \href{https://link.aps.org/doi/10.1103/PhysRevLett.51.1415}{Experimental
  {Tests} of the {``Invisible"} {Axion}}, Phys. Rev. Lett. 51 (1983)
  1415--1417.
\newblock \href {https://doi.org/10.1103/PhysRevLett.51.1415}
  {\path{doi:10.1103/PhysRevLett.51.1415}}.
\newline\urlprefix\url{https://link.aps.org/doi/10.1103/PhysRevLett.51.1415}

\bibitem{MADMAX}
P.~Brun, A.~Caldwell, L.~Chevalier, G.~Dvali, P.~Freire, E.~Garutti,
  S.~Heyminck, J.~Jochum, S.~Knirck, M.~Kramer, et~al., A new experimental
  approach to probe {QCD} axion dark matter in the mass range above
  40\,{\textmu}e{V}, Eur. Phys. J. C 79~(3) (2019) 1--16.

\bibitem{IAXO}
E.~Armengaud, F.~Avignone, M.~Betz, P.~Brax, P.~Brun, G.~Cantatore, J.~Carmona,
  G.~Carosi, F.~Caspers, S.~Caspi, et~al., Conceptual design of the
  international axion observatory {(IAXO)}, J. Instrum. 9~(05) (2014) T05002.

\bibitem{Bibber87}
K.~Van~Bibber, N.~Dagdeviren, S.~Koonin, A.~Kerman, H.~Nelson, Proposed
  experiment to produce and detect light pseudoscalars, Phys. Rev. Lett. 59~(7)
  (1987) 759.

\bibitem{alpstdr}
R.~B{\"a}hre, B.~D{\"o}brich, J.~Dreyling-Eschweiler, S.~Ghazaryan,
  R.~Hodajerdi, D.~Horns, F.~Januschek, E.-A. Knabbe, A.~Lindner, D.~Notz,
  et~al., Any light particle search {II}—technical design report, J. Instrum.
  8~(09) (2013) T09001.

\bibitem{Tanner}
P.~Sikivie, D.~Tanner, K.~van Bibber, Resonantly enhanced axion-photon
  regeneration, Phys. Rev. Lett. 98~(17) (2007) 172002.

\bibitem{Redondo:2010dp}
J.~Redondo, A.~Ringwald, Light shining through walls, Contemp. Phys. 52~(3)
  (2011) 211--236.

\bibitem{rc}
F.~Hoogeveen, T.~Ziegenhagen, Production and detection of light bosons using
  optical resonators, Nucl. Phys. 358~(1) (1991) 3--26.

\bibitem{Fukuda}
Y.~Fukuda, T.~Kohmoto, S.~I. Nakajima, M.~Kunitomo, Production and detection of
  axions by using optical resonators, Prog. Cryst. Growth Ch. 33~(1-3) (1996)
  363--366.

\bibitem{Gies}
H.~Gies, J.~Jaeckel, A.~Ringwald, Polarized light propagating in a magnetic
  field as a probe for millicharged fermions, Phys. Rev. Lett. 97~(14) (2006)
  140402.

\bibitem{DRD}
J.~H. P\~old, H.~Grote, {ALPS\,II} design requirements document (2019).

\bibitem{Magnet}
C.~Albrecht, S.~Barbanotti, H.~Hintz, K.~Jensch, R.~Klos, W.~Maschmann,
  O.~Sawlanski, M.~Stolper, D.~Trines, Straightening of superconducting {HERA}
  dipoles for the any-light-particle-search experiment {ALPS\,II}, EPJ
  Techniques and Instrumentation 8~(1) (2021) 5.

\bibitem{Arias}
P.~Arias, J.~Jaeckel, J.~Redondo, A.~Ringwald, Optimizing
  light-shining-through-a-wall experiments for axion and other weakly
  interacting slim particle searches, Phys. Rev. D 82~(11) (2010) 115018.

\bibitem{Bush}
Z.~R. Bush, S.~Barke, H.~Hollis, A.~D. Spector, A.~Hallal, G.~Messineo,
  D.~Tanner, G.~Mueller, Coherent detection of ultraweak electromagnetic
  fields, Phys. Rev. D 99~(2) (2019) 022001.

\bibitem{Jan}
J.~Dreyling-Eschweiler, N.~Bastidon, B.~D{\"o}brich, D.~Horns, F.~Januschek,
  A.~Lindner, Characterization, 1064\,nm photon signals and background events
  of a tungsten {TES} detector for the {ALPS} experiment, J. Mod. Optic.
  62~(14) (2015) 1132--1140.

\bibitem{HPLB}
T.~J. Kane, R.~L. Byer, Monolithic, unidirectional single-mode {Nd:YAG} ring
  laser, Opt. Lett. 10~(2) (2007) 65--67.

\bibitem{HPLC}
M.~Frede, B.~Schulz, R.~Wilhelm, P.~Kwee, F.~Seifert, B.~Willke, D.~Kracht,
  Fundamental mode, single-frequency laser amplifier for gravitational wave
  detectors, Opt. Express 15~(2) (2007) 459--465.

\bibitem{PDH}
R.~Drever, J.~L. Hall, F.~Kowalski, J.~Hough, G.~Ford, A.~Munley, H.~Ward,
  Laser phase and frequency stabilization using an optical resonator, Appl.
  Phys. B 31~(2) (1983) 97--105.

\bibitem{Black}
E.~D. Black, An introduction to {Pound--Drever--Hall} laser frequency
  stabilization, Am. J. Phys. 69~(1) (2001) 79--87.

\bibitem{Morrison}
E.~Morrison, B.~J. Meers, D.~I. Robertson, H.~Ward, Automatic alignment of
  optical interferometers, Appl. Opt. 33~(22) (1994) 5041--5049.

\bibitem{Buikema}
A.~Buikema, C.~Cahillane, G.~Mansell, C.~Blair, R.~Abbott, C.~Adams,
  R.~Adhikari, A.~Ananyeva, S.~Appert, K.~Arai, et~al., Sensitivity and
  performance of the {Advanced} {LIGO} detectors in the third observing run,
  Phys. Rev. D 102~(6) (2020) 062003.

\bibitem{Glover}
L.~Glover, M.~Goff, J.~Patel, I.~Pinto, M.~Principe, T.~Sadecki, R.~Savage,
  E.~Villarama, E.~Arriaga, E.~Barragan, et~al., Optical scattering
  measurements and implications on thermal noise in gravitational wave
  detectors test-mass coatings, Phys. Lett. 382~(33) (2018) 2259--2264.

\bibitem{Winkler}
W.~Winkler, K.~Danzmann, A.~R{\"u}diger, R.~Schilling, Heating by optical
  absorption and the performance of interferometric gravitational-wave
  detectors, Phys. Rev. A 44~(11) (1991) 7022.

\bibitem{Kwee:12}
P.~Kwee, C.~Bogan, K.~Danzmann, M.~Frede, H.~Kim, P.~King, J.~H. P{\~o}ld,
  O.~Puncken, R.~L. Savage, F.~Seifert, P.~Wessels, L.~Winkelmann, B.~Willke,
  \href{http://www.opticsexpress.org/abstract.cfm?URI=oe-20-10-10617}{Stabilized
  high-power laser system for the gravitational wave detector advanced {LIGO}},
  Opt. Express 20~(10) (2012) 10617--10634.
\newblock \href {https://doi.org/10.1364/OE.20.010617}
  {\path{doi:10.1364/OE.20.010617}}.
\newline\urlprefix\url{http://www.opticsexpress.org/abstract.cfm?URI=oe-20-10-10617}

\bibitem{Thorpe}
J.~I. Thorpe, K.~Numata, J.~Livas, Laser frequency stabilization and control
  through offset sideband locking to optical cavities, Opt. Express 16~(20)
  (2008) 15980--15990.

\bibitem{EPJTI}
J.~H. P{\~o}ld, A.~D. Spector, Demonstration of a length control system for
  {ALPS\,II} with a high finesse 9.2\,m cavity, Eur. Phys. J. TI 7~(1) (2020)
  1--9.

\bibitem{UFax}
G.~Mueller, P.~Sikivie, D.~Tanner, K.~Van~Bibber, Detailed design of a
  resonantly enhanced axion-photon regeneration experiment, Phys. Rev. D 80~(7)
  (2009) 072004.

\bibitem{REAPER}
G.~Mueller, P.~Sikivie, D.~B. Tanner, K.~Van~Bibber, Resonantly-enhanced
  axion-photon regeneration, in: AIP Conference Proceedings, Vol. 1274,
  American Institute of Physics, 2010, pp. 150--155.

\bibitem{BASEPLATE}
S.~Kulkarni, A.~Umi{\'n}ska, J.~Gleason, S.~Barke, R.~Ferguson, J.~Sanju{\'a}n,
  P.~Fulda, G.~Mueller, Ultrastable optical components using adjustable
  commercial mirror mounts anchored in a {ULE} spacer, Appl. Opt. 59~(23)
  (2020) 6999--7003.

\bibitem{Wei:20}
L.-W. Wei, K.~Karan, B.~Willke,
  \href{http://ao.osa.org/abstract.cfm?URI=ao-59-28-8839}{Optics mounting and
  alignment for the central optical bench of the dual cavity enhanced
  light-shining-through-a-wall experiment {ALPS\,II}}, Appl. Opt. 59~(28)
  (2020) 8839--8847.
\newblock \href {https://doi.org/10.1364/AO.401346}
  {\path{doi:10.1364/AO.401346}}.
\newline\urlprefix\url{http://ao.osa.org/abstract.cfm?URI=ao-59-28-8839}

\end{thebibliography}

\end{document}